\documentclass[12pt]{iopart}
\newcommand{\bbox}[1]{\mbox{\boldmath$#1$}}
\usepackage{iopams}
\usepackage{graphicx}

\renewcommand{\bbox}[1]{\mbox{\boldmath $#1$}}

\newcommand{\corr}[1]{\left\langle #1 \right\rangle}
\newcommand{\we}{\wedge}

\newcommand{\ints}[3]{\int_{#1}^{#2}{\rm d}#3}

\newcommand\nn{\nonumber}
\newcommand{\eqnlab}[1]{\label{eqn:#1}}

\newcommand{\lp}{\left(}
\newcommand{\rp}{\right)}

\newcommand{\lbp}{\left\{}
\newcommand{\rbp}{\right\}}
\newcommand{\ku}{\mbox{Ku}}

\newcommand{\DD}{{\cal D}}

\newcommand{\ve}[1]{\mbox{\boldmath$#1$}}

\newcommand{\ma}[1]{\mathbb{#1}}

\begin{document}

\title[]{Collisions of particles advected in random flows}

\author{K Gustavsson$^{(1)}$, B. Mehlig$^{(1)}$, and M. Wilkinson$^{(2)}$}

\address{
$^{(1)}$Department of Physics, G\"oteborg University, 41296
Gothenburg, Sweden\\
$^{(2)}$Department of Mathematics and Statistics,
The Open University, Walton Hall, Milton Keynes, MK7 6AA, England }
\begin{abstract}
We consider collisions of particles advected in a fluid. As already pointed
out by Smoluchowski [{\it Z. f. physik. Chemie}, {\bf XCII},
129-168, (1917)], macroscopic motion
of the fluid can significantly enhance the
frequency of collisions between the suspended particles.
This effect was invoked by Saffman and Turner
[{\it J. Fluid Mech.}, {\bf 1}, 16-30, (1956)]
to estimate collision rates of small water droplets in turbulent rain clouds,
the macroscopic motion being caused by turbulence.
Here we show that the Saffman-Turner theory is
unsatisfactory because it describes
an initial transient only. The reason for this failure is that the
local flow in the vicinity of a particle is treated as if
it were a steady hyperbolic flow, whereas in reality it must fluctuate.
We derive exact expressions for the steady-state collision
rate for particles suspended in rapidly fluctuating
random flows and compute
how this steady state is approached.
For incompressible flows, the Saffman-Turner expression
is an upper bound.
\end{abstract}

\maketitle
\newpage


\newpage
\section{Introduction}
\label{sec: 1}

Turbulent aerosols are of interest in a variety of natural
and technological systems. Two very important examples are water droplets in turbulent rain clouds \cite{Sha03} and dust grains in turbulent accretion disks around growing stars \cite{Wil08}. In both of these systems the aerosol is unstable because the suspended particles collide (leading to aggregation or possibly fragmentation of the aerosol particles). The collision processes therefore have significant consequences: the formation of rain in one case, and the widely hypothesised mechanism for the formation of planets in the other. Collisions always occur due to molecular diffusion, but (as pointed out by Smoluchowski
\cite{Smo17,Saf56}),
macroscopic fluid motion can considerably increase the collision rate.

If the suspended particles are sufficiently heavy (so that their inertia becomes relevant),
they can move relative to the fluid. In this case the occurrence of \lq caustics' will typically
increase the collision rate by several orders of magnitude \cite{Fal02,Wil06}. If the aerosol particles are sufficiently
light, their \lq molecular' diffusion can make a significant contribution
to the collision rate, which can be estimated by standard kinetic theory.

In this paper we are concerned with the effect of macroscopic motion
of a fluid on small particles which have insignificant inertia, so that
they follow the flow (advective motion). The seminal papers in this
area were due to Smoluchowski \cite{Smo17}, who first considered the effect
of shear of the fluid flow on collisions, and Saffman \& Turner \cite{Saf56}, who
gave a formula for the collision rate which has formed the basis
for most subsequent work on this problem. Their paper was motivated
by a problem in meteorology argued that small-scale turbulence in convecting clouds can accelerate collisions between microscopic water droplets, thus initiating rain formation:  in this case the particles are brought into contact by hyperbolic or shearing motions of the turbulent flow.

The formula for the collision rate by Saffman \& Turner \cite{Saf56} has been used frequently in the past five decades in cloud physics and in chemical engineering problems. It appears to be widely accepted that
their expression is an exact relation for the collision rate in a dilute
suspension. In the following we show that the Saffman-Turner estimate describes an initial transient of the problem only. For particles suspended in incompressible flows, the collision rate falls below the initial transient (which thus constitutes an upper bound). For particles advected in a compressible flows, however, homogeneously distributed particles will cluster in a compressible fluid (see for example \cite{Bal01,Fal01}).
The clustering may increase the collision rate beyond its initial transient.

The Saffman-Turner approximation treats the flow surrounding a test particle as if it were a steady hyperbolic flow, while in reality the flow fluctuates as a function of time. In section \ref{sec: 3.2} below we give an extension of the Saffman-Turner formula which does give the collision rate exactly.
Unfortunately the formula contains information about the time-dependence
of the flow, and it is impossible to evaluate it in the general case.

Because of the importance of understanding collision rates for aerosol particles, it is desirable to find exactly solvable cases which can be
used as a benchmark for numerical studies.
The collision rate must depend on a dimensionless parameter describing how quickly the fluid velocity fluctuates, the \lq Kubo number' \cite{Wil07}. We are able to obtain precise asymptotic results on the collision rate in the limit where the Kubo number approaches zero. In this case, particle separations undergo a diffusion process. By solving the corresponding Fokker-Planck equation we can determine the collision rate exactly.

The remainder of this paper is organised as follows. In section \ref{sec: 2} we introduce the equations of motion and the dimensionless parameters
of the problem. Section \ref{sec: 3} discusses the Saffman-Turner
theory and our extension of it. Section \ref{sec: 3.1} describes the expression for the collision rate given in \cite{Saf56} which is the starting point of our discussions. (Some new results on the
evaluation of the Saffman-Turner expression are described in the appendix). In section \ref{sec: 3.2} we discuss our exact formula for the collision rate, and explain why the Saffman-Turner approximation describes an initial transient only. In sections \ref{sec: 4}-\ref{sec: 6} we discuss how exact asymptotic results may be found for the limit of small Kubo number. A Fokker-Planck equation for the probability density of the separation of particles is described in section \ref{sec: 4}. In section \ref{sec: 5} this is used to obtain the steady-state collision rate and section \ref{sec: 6} gives the full time dependence of the collision rate. These results are compared to numerical simulations in section \ref{sec: 7}, which also contains some concluding remarks, discussing scope for further work in this area.

Finally, we remark that a brief summary of some of the results of this paper has already been published \cite{And07}. Here we discuss the problem in greater depth and generality, and derive expressions for the time-dependent collision rate which were not discussed in \cite{And07}.


\section{Equations of motion and dimensionless variables}
\label{sec: 2}

We consider spherical particles of radius $a$ in a fluid with velocity field  $\ve u(\ve r,t)$ which has an apparently random motion, usually as a result of turbulence. We assume that the suspended particles do not modify the surrounding flow. When the inertia of the particles is negligible, they are advected by the flow:
\begin{equation}
\label{eq: 2.1}
\dot{\ve r}=\ve u(\ve r,t)\,.
\end{equation}
It is assumed that direct interactions between the particles can be neglected until they collide. In other words, the particles follow equation
(\ref{eq: 2.1}) until their separation falls below $2a$.

We model the complex flow of a turbulent fluid by a random velocity field $\mbox{\boldmath$u$}(\mbox{\boldmath$r$},t)$. We consider flows in both two and three spatial dimensions and for convenience we use a Gaussian distributed field when we carry through concrete computations. In most cases we are concerned with incompressible flow, satisfying
$\mbox{\boldmath$\nabla$}\cdot \mbox{\boldmath$u$}=0$. Particles floating
on the surface of a fluid may experience a partly compressible flow \cite{Cre04,Fal05}, as may particles in gases moving with speeds comparable to the speed of sound. For these reasons we also consider partially compressible flows.

It is convenient to construct the random velocity field $\ve u(\ve r,t)$ from scalar stream functions or potentials \cite{Wil07}. In two spatial dimensions we write
\begin{equation}
\label{eq: 2.2}
\bbox{u} ={\cal N}_2\,
(\bbox{\nabla}\wedge \psi \hat {\bf n}_z +
\beta\bbox{\nabla}\phi)
\end{equation}
where ${\cal N}_2$ is a normalisation factor and $\psi$ and $\phi$
are independent Gaussian random functions. We shall use angle brackets to denote averaging throughout. The fields $\phi(\mbox{\boldmath$r$},t)$ and $\psi(\mbox{\boldmath$r$},t)$ have zero averages, $\langle \phi\rangle=0$,
$\langle\psi\rangle=0$ and they both have same correlation function, $C(R,T)$:
\begin{equation}
\label{eq: 2.3}
\langle \phi(\mbox{\boldmath$r$},t)\phi(\mbox{\boldmath$r$}',t')\rangle=C(R,T)
\end{equation}
where $R = |\ve r-\ve r'|$ and $T = |t-t'|$. This two-point correlation function $C(R,T)$ is a smooth function decaying (sufficiently rapidly) to zero for large values of $R$ and $T$. For $\beta = 0$, the flow (\ref{eq: 2.2}) is incompressible. For finite values of $\beta$ it acquires a compressible component. In the limit of $\beta\rightarrow \infty$ the flow is purely potential. In some cases the physics of a problem dictates that $\psi$ and $\phi$ should have different correlation functions; many of our results can be generalised in this way.

In three spatial dimensions we write
\begin{equation}
\label{eq: 2.4}
\ve u={\cal N}_3\,\left(\ve\nabla\we\ve
A+\beta\ve\nabla\phi\right)
\end{equation}
where $\ve A = (A_1, A_2, A_3)$ and $\phi$ are four independent scalar fields with zero mean and with the same correlation function $C(R,T)$; and ${\cal N}_3$ is a normalisation factor.

In the remainder of this paper we choose the normalisation factors to be
of the form
\begin{equation}
\label{eq: 2.5}
{\cal N}_d =\frac{u_0}{\sqrt{-d(d-1+\beta^2)C''(0,0)}}
\end{equation}
where $u_0$ is the standard deviation of the magnitude of the velocity
and $C''$ denotes the second derivative of the correlation function
(\ref{eq: 2.3}) with respect to its first argument.

Fully-developed turbulent flows have a power-law spectrum in the inertial range, covering a wide band of wavenumbers \cite{Fri97,Fal01}. This feature can be incorporated by giving $C(R,T)$ a suitable algebraic behaviour over a range of values of $R$, as explained in \cite{And07}. The long-ranged behaviour of the velocity field is not, however, relevant to the advective collision mechanism. It therefore suffices to consider a model with a short-ranged velocity correlation: for the numerical work reported in this paper we used following form of the correlation function
\begin{equation}
\label{eq: 2.6}
C(R,T)=C_0 \exp\Big({-\frac{R^2}{2\eta^2}-\frac{T}{\tau}}\Big)
\end{equation}
where $C_0$ is a constant. In our numerical simulations we represent the flow field by its Fourier components which are subject to an Ornstein-Uhlenbeck process as suggested by Sigurgeirson \& Stuart \cite{Stu02}.

\begin{table}
\begin{tabular}{lc}
\hline\\[-5mm]\hline
\mbox{Parameter} & \mbox{Symbol}\\
\hline
Particle size &$a$\\
Typical velocity fluctuation & $u_0$\\
Correlation length of the flow & $\eta$\\
Correlation time of the flow & $\tau$\\
Number density of particles & $n_0$ \\
Compressibility & $\beta$\\
Spatial dimension & $d$\\
\hline\\[-5mm]\hline
\end{tabular}
\vspace*{3mm}
\caption{\label{tab:1} Parameters of the model.}
\end{table}

Our problem is characterised by the six parameters listed in table \ref{tab:1}. From the parameters in table \ref{tab:1}, three independent dimensionless combinations can be formed:
\begin{equation}
\label{eq: 2.7}
{\rm Ku} = u_0 \tau/\eta\,,\quad n_0 a^d\,,\quad \bar a= 2a/\eta\,.
\end{equation}
The first parameter characterises the dimensionless speed of the flow and is called Kubo number, discussed in \cite{Wil07}. Note that $\ku\gg 1 $ is not possible, because the motion of the fluid places an upper limit on the correlation time. Steady, fully  developed turbulence corresponds to  $\ku\sim 1$. The Kubo number can be small for randomly stirred fluids.
It is only in the limit ${\rm Ku}\to 0$ that we are able to obtain precise
and explicit estimates for the collision rate: this case is considered
in sections \ref{sec: 4}-\ref{sec: 6}. The second parameter in (\ref{eq: 2.7}) is the packing fraction of particles. Throughout we assume that this parameter is small. Similarly, the third parameter in (\ref{eq: 2.7}) is usually taken to be small.

We note that Kalda \cite{Kal07} has considered the collision rate for particles in a non-smooth velocity field: this could be relevant to the case where $a\gg \eta$.


\section{Evaluating the collision rate}
\label{sec: 3}

\subsection{The Saffman-Turner expression}
\label{sec: 3.1}

Throughout this paper we consider the rate of collision of a given
particle with any other particle, denoting this by ${\cal R}$. Some papers
consider the total rate of collision per unit volume. If the spatial
density of particles is $n_0$, the total rate of collision per unit
volume is $\frac{1}{2}n_0{\cal R}$ (the factor of $\frac{1}{2}$ avoids double-counting). We will not be concerned with what happens
after particles undergo their first collision: in different physical
circumstances they may coalesce, scatter, or fragment, but in this
paper we are concerned only with their first contact.

We assume that the particles are spherical (or circular, in two-dimensional
calculations) and that they all have the same radius, $a$. We regard
the particles as having collided when their separation reaches $2a$,
and we neglect effects due to the interaction of the particles and the
fluid. In practice the fluid trapped between approaching particles may
slightly reduce the collision rate \cite{Saf56}, but this effect can be accounted for by replacing the radius by an effective radius.

The problem of calculating the rate of collision therefore reduces to
the following problem. We consider a given particle, and transform to
a frame where the centre  of this particle is at the origin, and the
separation of the centre of another particle is denoted by $\mbox{\boldmath$R$}$. Initially, the reference particle is surrounded by a gas
of particles with spatial density $n_0$. We assume that these are initially randomly distributed, apart from the constraint that none of the particles is in contact with the reference particle. Collisions with the test particle occur when other particles come within a radius $2a$ of the reference particle. The rate of collisions is therefore the rate at which particles cross a sphere of radius $2a$ centred at the origin in the relative coordinate system. This is obtained by integrating the inward radial
velocity over the sphere, and multiplying by the density $n_0$. This approach gives an expression for the collision rate which we term ${\cal R}_0$:
\begin{equation}
\label{eq: 3.1}
{\cal R}_0= -n_0\ints{}{}{\Omega} \,
v_r(2a,\Omega,t)\,\Theta(-v_r(2a,\Omega,t))\,.
\end{equation}
where $v_r(R,\Omega,t)$ is the radial velocity at spherical coordinate
$\Omega$, radius $R$ and time $t$. The function $\Theta(x)$ is a Heaviside step function, which is used to select regions of the surface where the flow is into the sphere of radius $2a$. This is the fundamental expression for the collision rate given by Saffman \& Turner \cite{Saf56}. In section
\ref{sec: 3.2} we discuss why this expression is not exact, and give the precise formula. The remainder of this section considers how this expression is evaluated; some new results are presented in the appendix.

\begin{figure}[t]
\begin{center}
\includegraphics[width=6cm,clip]{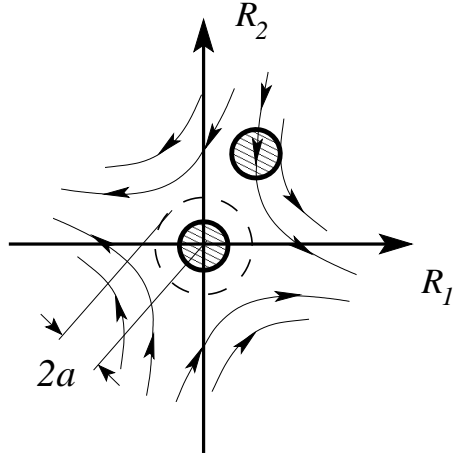}
\caption{\label{fig: 1}
Schematic picture of two particles of radius $a$ passing each other in a hyperbolic flow. The particle at the origin will see particles move past on hyperbolic trajectories. Collisions occur whenever particles approach closer than $2a$. The collision rate is thus determined by
the influx of particles into a disc of radius $2a$ (dashed line)
around the origin.}
\end{center}
\end{figure}

The evaluation of (\ref{eq: 3.1}) is greatly simplified in the case
where the particles are small, in the sense that $a/\eta \ll 1$. In this case the relative velocity is accurately approximated using the velocity gradient of the random velocity field $\mbox{\boldmath$u$}$. This approximation was also considered by Saffman \& Turner \cite{Saf56}. To lowest order in ${R}$, the relative velocity $\ve{\dot R}=\ve{u}(\ve{r}+\ve{R},t)-\ve{u}(\ve{r},t)$ is approximated as $\ma{A}(\ve{r},t)\ve{R}$ where $\ma{A}$ is the rate of strain matrix of the flow $\ve{u}$, with elements $A_{ij}={\partial u_i}/{\partial r_j}$ (and $i,j = 1,\ldots,d$). Two particles with radii $a\ll\eta$ moving close to each other will thus experience a relative velocity $\ma{A}_0\ve{R}$. This is illustrated in figure \ref{fig: 1} for a flow which is hyperbolic in the vicinity of the reference particle.
The corresponding relative radial speed $v_r$ is
$v_r=\ve{\hat n}^T\ma{A}_0\ve{R}$ where $\ve{\hat n}$ is the radial unit vector. In particular, at the distance $R=2a$ where particles collide, the radial speed is $v_r=2a\ve{\hat n}^T\ma{A}_0\ve{\hat n}$. Thus when $a/\eta \ll 1$, equation (\ref{eq: 3.1}) reduces to
\begin{equation}
\label{eq: 3.2}
{\cal R}_0
=  -2an_0\ints{}{}{\Omega}\,\ve{\hat n}^T\ma{A}_0\ve{\hat n}\,
\Theta(-\ve{\hat n}^T\ma{A}_0\ve{\hat n})\,.
\end{equation}
This expression was also obtained in \cite{Saf56}.

The remainder of this subsection is concerned with the evaluation of (\ref{eq: 3.2}).
Earlier, Smoluchowski \cite{Smo17} had considered the special case of a fluid flowing with a uniform shear in $3$ dimensions:
\begin{equation}
\label{eq: 3.3}
\ve u =
\left(\begin{array}{c}
\alpha z\\
0\\
0
\end{array} \right)\,,
\quad
\ma A_0 =
\left(\begin{array}{ccc}
0&0 &\alpha\\
0&0 &0\\
0&0 &0
\end{array} \right)\,.
\end{equation}
He obtained the collision rate:
\begin{equation}
\label{eq: 3.4}
{\cal R}_0=\frac{4}{3}\alpha n_0 (2a)^3
\end{equation}
which is in agreement with the result of evaluating (\ref{eq: 3.2}) for this case. For a general strain-rate matrix ${\ma A}_0$ the evaluation of
(\ref{eq: 3.2}) is, however, very difficult. In \ref{sec: appA} we discuss how this expression is evaluated for a general matrix $\ma A_0$
in two dimensions, and for a general traceless matrix (representing an
incompressible flow) in three dimensions.

In a turbulent flow, Saffman \& Turner \cite{Saf56} argued that the elements of $\ma{A}_0$ change as a function of position and one needs to average over the ensemble of strain matrices ${\ma A}_0$ at different positions in order to estimate the collision rate, so that (\ref{eq: 3.2}) is replaced by:
\begin{equation}
\label{eq: 3.5}
{\cal R}_{0} =  -2an_0\Big\langle \ints{}{}{\Omega} \,\,
\ve{\hat n}^T\ma{A}_0\ve{\hat n}\ \Theta(-\ve{\hat n}^T\ma{A}_0\ve{\hat n})\Big\rangle\,.
\end{equation}
At first sight the requirement to average over $\ma A_0$ appears to
complicate the problem. However, for a rotationally invariant ensemble
of random flows, the problem is considerably simplified by taking
the average. In an incompressible flow, for each realisation of $\ve u(\ve r,t)$, the  currents into and out of the collision region (the disk or
sphere of radius $2a$) cancel precisely. Therefore (\ref{eq: 3.5}) can
be written as
\begin{equation}
\label{eq: 3.6}
{\cal R}_{0} =  an_0 \Big\langle \ints{}{}{\Omega} \, \,
|\ve{\hat n}^T\ma{A}_0\ve{\hat n}|\Big\rangle\,.
\end{equation}
The same result also holds for cases where the flow is compressible,
of the form (\ref{eq: 2.2}) or (\ref{eq: 2.4}), with $\beta\ne 0$. This is demonstrated by the following argument (here we discuss the two-dimensional case). If we did not include the factor
$\Theta(-\ve{\hat n}^T\ma{A}_0\ve{\hat n})$ in (\ref{eq: 3.6}), we would calculate the sum
of the collision rate for the flow $\mbox{\boldmath$u$}$ and for the
time-reversed flow $-\mbox{\boldmath$u$}$. The time reversed flow is generated by reversing the signs of the potentials $\phi$ and $\psi$ in
(\ref{eq: 2.2}). The probability density for the Gaussian field
$-\phi$ is the same as for $\phi$ (and similarly for $\psi$). It
follows that the collision rate for the time-reversed flow
$-\mbox{\boldmath$u$}$ is the same as for $\mbox{\boldmath$u$}$.
The expression (\ref{eq: 3.6}) is therefore also valid for the
cases of compressible flow which we consider in this paper.

Now using rotational symmetry, one finds the very simple expression
\begin{equation}
\label{eq: 3.7}
{\cal R}_{0} =  an_0 A_d(2a) \langle |A_{11}|\rangle
\end{equation}
where $A_d(2a)$ is the the area of the sphere of radius $2a$ in $d$
dimensions (explicitly $A_2(r)=\pi r^2$ and $A_3(r)=4\pi r^3/3$). For the Gaussian model flow which we consider, equation (\ref{eq: 3.7}) gives
\begin{equation}
\label{eq: 3.8}
{\cal R}_{0} =
\frac{n_0}{\sqrt{2\pi}}A_d(2a)2a\sqrt{\corr{A_{11}A_{11}}}
\end{equation}
with
\begin{equation}
\label{eq: 3.9}
\corr{A_{11}A_{11}}=-\frac{{\cal N}_d^2(d-1+3\beta^2)C''''(0,0)}{3}\,.
\end{equation}
%
%


\subsection{An exact expression for the collision rate}
\label{sec: 3.2}

The Saffman-Turner expression for the collision rate, equation (\ref{eq: 3.1}) or (\ref{eq: 3.7}), correctly describes the initial collision rate. There are, however, two reasons why the collision rate may approach a significantly different value after an initial transient.

\begin{figure}[t]
\begin{center}
\includegraphics[width=15cm]{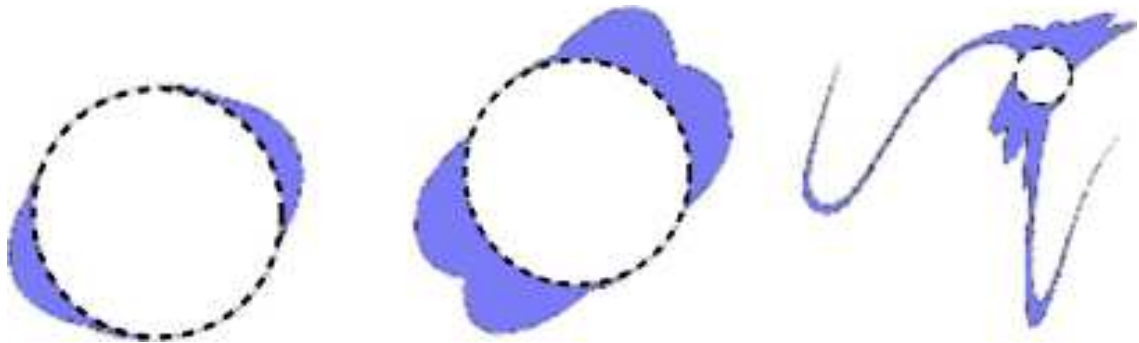}
\caption{\label{fig: 2} Initial positions of particles which
have collided with a reference particle at the origin after a certain time
(shaded regions).
The collision region (disk of radius $2a$) is bounded by a dashed circle.
At short times this region extends along the direction of the stable eigenvector of the initial flow. At intermediate times, we see the effect of this eigenvector having rotated. At long times, the set of colliding initial conditions is stretched and folded. The shaded region then covers most of the surface of the circle, including the region where the radial velocity is inward.
The figure shows results of a simulated flow with small Kubo number, of the type discussed in section \ref{sec: 4}, at times $t=20\tau$, $t=100\tau$ and $t=1000\tau$. }
\end{center}
\end{figure}

The first reason why (\ref{eq: 3.1}) may fail arises from the fact that the flow field fluctuates in time. Recall that we are concerned with the rate
at which pairs of particles collide for the first time. If the flow is time-dependent, the relative position coordinate $\mbox{\boldmath$R$}(t)$
may pass through the sphere of radius $2a$ more than once. This effect may be accounted for by writing the collision rate in the form
\begin{equation}
\label{eq: 3.13}
{\cal R}(t)
= -n(t)\ints{}{}{\Omega} \,
v_r(2a,\Omega,t)\,\Theta(-v_r(2a,\Omega,t))
\chi(2a,\Omega,t)\,.
\end{equation}
As before ${\rm d}\Omega$ is the $d$-dimensional surface element at $R=2a$
and $v_r$ is the radial velocity component (the relative speed) and the Heaviside step function ensures that only particles entering the sphere contribute to the collision rate. The factor $n(t)$ is the density of particles in the neighbourhood of the test particle at time $t$. The function $\chi$ is an indicator function: it is equal to unity if the point reaching the surface element $\Omega$ at radius $2a$ and at time $t$ has not previously passed through the sphere of radius $2a$, otherwise it is zero.
The effect of including the function $\chi$ is illustrated
in figure \ref{fig: 2}. The collision rate will reduce below the Saffman-Turner estimate for times where $\chi$ is no longer unity..

In the general case (\ref{eq: 3.13}) is difficult to evaluate since the
indicator function $\chi$ depends on the history of the flow. If the flow is rapidly fluctuating (that is, if the Kubo number of the flow is small), the relative separation of two particles undergoes a diffusion process which makes it possible to exactly evaluate the collision rate for a given ensemble of random flows.

A second effect may modify the collision rate from the Saffman-Turner estimate is that if the flow field is compressible, particles may cluster together, and the particle density in the vicinity of a test particle may be higher than the average density. This effect is expected to cause the collision rate to increase, after an initial transient during which the clustering becomes established. Again, this effect is very hard to quantify in the general case, but we can obtain precise asymptotic results in the limit of small Kubo number, where a diffusion approximation becomes applicable. Because we consider particles which are advected by the flow, there is no clustering effect when the flow is incompressible.

We note that for incompressible flow the only correction to the Saffman-Turner estimate is the occurrence of the factor $\chi$ in
(\ref{eq: 3.13}). This implies that for advected particles in an incompressible flow, the Saffman-Turner estimate of the collision rate is an upper bound.


\section{Diffusion approximation for small Kubo number}
\label{sec: 4}

In the limit of a rapidly changing flow, that is when $\ku \ll 1$,  particle separations $\ve R = \ve r -\ve r'$ undergo a diffusion process  (see \cite{Fal01} for a review). By solving the corresponding Fokker-Planck equation with the appropriate boundary conditions, we can determine the collision rate exactly in this limit.

\subsection{Fokker-Planck equation}
\label{sec: 4.1}

Consider two particles, one at $\ve r$ and the other in its vicinity, at $\ve r'$. The equation of motion (\ref{eq: 2.1}) implies that their separation
$\ve R=\ve r'-\ve r$ obeys
\begin{equation}
\dot{\ve R}=\ve{u}(\ve r+\ve R,t)-\ve{u}(\ve r,t)\,.
\label{eq: 4.1}
\end{equation}
Integrating (\ref{eq: 4.1}) over a short time interval $\delta t$ we obtain
\begin{equation}
\delta{\ve R}(\ve r,t) = \ints{t}{t+\delta t}{t'}(\ve u(\ve r+\ve R,t')-\ve u(\ve r,t'))\,.
\end{equation}
The moments of the components $\delta R_i$ of $\delta \ve R$ are:
\begin{eqnarray}
\corr{\delta{R_i}} &=& \ints{t}{t+\delta t}{t'}\corr{u_i(\ve r+\ve R,t')-u_i(\ve r,t')}=0\nn\\
\corr{\delta{R_i}\delta{R_j}}
&\approx& \delta t\ints{-\infty}{\infty}{t}\corr{(u_i(\ve R,t)-u_i(\ve 0,t))(u_j(\ve R,0)-u_j(\ve 0,0))}\,,
\end{eqnarray}
assuming that $\delta t \gg \tau$. We thus obtain the following Fokker-Planck equation for the density $\rho(\ve R,t)$ of particle separations
\cite{Bal01,Fal01,And07}:
\begin{equation}
\frac{\partial\rho}{\partial t} = {\ve \nabla}^T {\ma D}({\ve R}) {\ve \nabla} \rho\,.
\label{eq: 4.4}
\end{equation}
Here ${\ma D}({\ve R})$ is a diffusion matrix with diffusion coefficients
\begin{eqnarray}
\label{eq: 4.5}
D_{ij}(\ve R)&=\frac{1}{2\delta t}\corr{\delta{R_i}\delta{R_j}} \nn\\
&=\ints{-\infty}{\infty}{t}\left[\corr{u_i(\ve 0,t)u_j(\ve 0,0)}-\corr{u_i(\ve R,t)u_j(\ve 0,0)}\right]\,.
\end{eqnarray}
In terms of the correlation function $C(R,T)$ introduced in section
\ref{sec: 2} we have
\begin{equation}
\label{eq: 4.6}
\begin{array}{ll}
D_{ij}(\ve R)&= -{\cal N}_d^2
\int_{-\infty}^{\infty}{\rm d}t\bigg[\delta_{ij}\Big((d-1+\beta^2)C''(0,t)\nn\\
&\hspace*{2cm}-\frac{d-2+\beta^2}{R}C'(R,t)-C''(R,t)\Big)\cr
&\hspace*{2cm}+(1-\beta^2)\frac{R_iR_j}{R^2}\lp-\frac{1}{R}
C'(R,t)+C''(R,t)\rp\bigg]\,.
\end{array}
\end{equation}
In order to determine the collision rate it is convenient to write the Fokker-Planck equation in the form of a continuity equation
\begin{equation}
\label{eq: 4.7}
\frac{\partial\rho}{\partial t}+\ve\nabla^T\ve j=0
\end{equation}
where the components of the probability current can be identified as
$\ve j = -\ma D(\ve R)\ve \nabla \rho$. The collision rate ${\cal R}(t)$
between particles of radius $a$ is given by the rate at which the particle separation $R$ decreases below $2a$ \cite{And07}. This rate equals the radial probability current of particle separations evaluated at $R=2a$, i.e.
\begin{equation}
{\cal R}(t)=-\int{\rm d}\Omega\ \hat{\ve n}^T{\ve j}(R=2a,t)
\label{eq: 4.8}
\end{equation}
where $\hat{\ve n}$ is the radial unit vector. We evaluate (\ref{eq: 4.8}) in two steps. First (in section \ref{sec: 5}) we determine the steady-state collision rate as $t\rightarrow \infty$. Second, we determine the full time-dependence of the collision rate (section \ref{sec: 6}), describing how the initial transient discussed in section \ref{sec: 3} approaches the steady state.

\subsection{Transformation to spherical coordinates}
\label{sec: 4.2}

Because of the angular symmetry of the statistics of the fluid velocity, it is convenient to transform (\ref{eq: 4.4}) to spherical coordinates. The probability density $\rho$ depends on the separation $R$ only and obeys
\begin{equation}
R^{d-1}\frac{\partial\rho}{\partial t}(R,t)= -\frac{\partial}{\partial R}\lp R^{d-1} j_r(R,t)\rp\,.
\label{eq: 4.9}
\end{equation}
Here the radial probability current $j_r(R,t)$ is given by
\begin{equation}
j_r(R,t)=-\beta^2{\cal N}_d^2f(R)\rho(R,t)+{\cal N}_d^2g(R)\frac{\partial}{\partial R}\rho(R,t)\,,
\label{eq: 4.10}
\end{equation}
with
\begin{eqnarray}
\label{eq: 4.11}
f(R)&=\ints{-\infty}{\infty}{t}
\Big[\frac{1-d}{R}\lp\frac{1}{R} C'(R,t)-C''(R,t)\rp+C'''(R,t)\Big]\,,\nn\\
g(R)&=\ints{-\infty}{\infty}{t}\Big[(d-1+\beta^2)C''(0,t)
-\frac{d-1}{R}C'(R,t)-\beta^2C''(R,t)\Big]\,.
\end{eqnarray}
%
%


\section{Steady-state solution}
\label{sec: 5}

Consider now the steady-state solution of
(\ref{eq: 4.4}), obeying $\partial\rho/\partial t = 0$.
The steady-state collision rate, denoted by ${\cal R}_\infty$, is determined by (\ref{eq: 4.8}) in terms of the steady-state current. The total current $-A_d(R)j_r(R)$ entering a sphere of radius $R$ and area $A_d(R)$ must be a constant. This gives $j_r(R)=-{\cal R}_\infty/A_d(R)$ where ${\cal R}_\infty$ is a constant, equal to the collision rate which we wish to determine.

The equation of motion (\ref{eq: 2.1}) is only physically meaningful when $a\ll\eta$, so that the fluid velocity is approximately the same throughout all of the region occupied by the particle. We will, however, solve the diffusion equation for general $a$. One reason for treating the general case is that it is hard to verify the analytical expressions for the limiting case $a/\eta\ll1$ in numerical investigations.

To determine this current, it is necessary to consider the appropriate
boundary conditions. First, if the particle distribution initially is uniform with density $n_0$, we expect it to remain so at large separations. Thus we use the boundary condition
\begin{equation}
\rho(R\rightarrow\infty,t)=n_0\,.
\label{eq: 5.1}
\end{equation}
The boundary condition (\ref{eq: 5.1}) is implemented as follows. For a specified current density $j_r(R,t)$ we can solve (\ref{eq: 4.10}) by finding an integrating factor $h$:
\begin{equation}
\label{eq: 5.2}
\rho(R,t)=h(R)\lp A-\ints{R}{\infty}{R'}\frac{j_r(R',t)}{g(R')h(R')}\rp
\end{equation}
with  integration constant $A$ and
\begin{equation}
\label{eq: 5.3}
h(R)=\exp\lp\ints{R}{\infty}{R'}\frac{f(R')}{g(R')}\rp\,.
\end{equation}
The boundary condition (\ref{eq: 5.1})  determines the integration constant in (\ref{eq: 5.2}) to be $A_0$. Second, when particles comes closer than the distance of $2a$ they collide and must be removed. This is taken care of by the following boundary condition at $2a$:
 \begin{equation}
\rho(2a,t)=0.
\label{eq: 5.4}
\end{equation}
Inserting this boundary condition  into (\ref{eq: 5.2}) yields
\begin{equation}
\label{eq: 5.5}
n_0 = \int_{2a}^\infty\!{\rm d}R' \frac{j_r(R')}{g(R') h(R')}\,.
\end{equation}
Using $j_r(R') =- {\cal R}_\infty/A_d(R')$ and solving for the constant ${\cal R}_\infty$ we find the following expression for the collision rate:
\begin{equation}
\label{eq: 5.6}
{\cal R}_\infty=-\frac{2\pi^{d/2} n_0}{\Gamma(d/2)}{\cal N}_d^2\Bigg[\ints{2a}{\infty}{R}\frac{R^{1-d}}{g(R)}
\exp\lp\beta^2\ints{R}{\infty}{R'} \frac{f(R')}{g(R')}\rp
\Bigg]^{-1}
\end{equation}
where $\Gamma(z)$ is the Gamma function. Equation (\ref{eq: 5.6})
is the main result of this section. We continue by discussing a number of limiting cases.

\subsection{Collision rate for incompressible flows}
\label{sec: 5.1}

For an incompressible flow ($\beta=0$), the integrating factor is just unity
(incompressibility $\ve \nabla^T \ve u=0$ implies $\ve\nabla^T\ma D ={\mathbf 0}$
which in turn implies that the term proportional to $\rho$ in
(\ref{eq: 4.10}) vanishes). In this case, equation (\ref{eq: 5.6}) simplifies to
\begin{eqnarray}
{\cal R}_\infty&=\frac{2\pi^{d/2}(2a)^{d-1}}{\Gamma(d/2)}\frac{n_0u_0^2}{dC''(0,0)}\nn\\
&\hspace*{1cm}\times\left[\ints{2a}{\infty}{R}
\left({\ints{-\infty}{\infty}{t}\Big(
R^{d-1}C''(0,t)-R^{d-2}C'(R,t)\Big)}\right)^{-1}\right]^{-1}\hspace*{-3mm}\,.
\label{eq: 5.7}
\end{eqnarray}

\subsection{Collision rate for $\beta = 1$}
\label{sec: 5.2}

When $\beta=1$, the strength of the solenoidal and potential parts of the field are equal. This case may be relevant to the dynamics of particles floating on the surface of a turbulent fluid \cite{Cre04}. The collision rate can be evaluated exactly in this case, by rewriting (\ref{eq: 5.6}) as
\begin{eqnarray}
{\cal R}_\infty
&=\frac{2\pi^{d/2}}{\Gamma(d/2)}\frac{n_0u_0^2}{d(d-1+\beta^2)C''(0,0)}{\rm e}^{\beta^2\ln g(\infty)}\nn\\
&\times
\Bigg[\!\int_{2a}^{\infty}\!\!\!\!{\rm d}{R}R^{1-d}g(R)^{\beta^2-1}\exp
\Big(\beta^2(\beta^2\!-\!1)
\int_{R}^{\infty}\!\!\!\!{\rm d}{R'}
\frac{\ints{-\infty}{\infty}{t}C'''(R',t)}{g(R')}\Big)
\Bigg]^{-1}\,.
\label{eq: 5.8}
\end{eqnarray}
Setting $\beta =1$ we find
\begin{eqnarray}
{\cal R}_\infty=&
\frac{2\pi^{d/2}}{\Gamma(d/2)}\frac{n_0u_0^2}{d^2C''(0,0)}&\nonumber\\
&\times
\frac{\ints{-\infty}{\infty}{t}\lp dC''(0,t)-\lim_{R'\rightarrow\infty}\frac{d-1}{R'}C'(R',t)-C''(\infty,t)\rp}{\ints{2a}{\infty}{R}R^{1-d}}&
\label{eq: 5.9}
\end{eqnarray}
for $d\leq 2$. When  $d>2$ we obtain
\begin{eqnarray}
{\cal R}_\infty=&
(d-2)(2a)^{d-2}\frac{2\pi^{d/2}}{\Gamma(d/2)}\frac{n_0u_0^2}{d^2C''(0,0)}&\nonumber\\
&\times
\ints{-\infty}{\infty}{t}\lp dC''(0,t)-\lim_{R'\rightarrow\infty}\frac{d-1}{R'}C'(R',t)-C''(\infty,t)\rp\,.&
\label{eq: 5.10}
\end{eqnarray}
What makes it possible to find exact solutions in this case is the fact that  when $\beta=1$,
the functions $f(R)$ and $g(R)$ in (\ref{eq: 4.11}) are related as $g'(R)=f(R)$.
Note that this relation is always true in one spatial dimension, independent of the value of $\beta$.
Note also that when $g'(R)=f(R)$,
equation  (\ref{eq: 4.10})  can be solved for $\rho$
by integration (starting at e.g. $2a$, with $\rho(2a,t)=0$)
\begin{eqnarray}
\rho(R)&=\frac{1}{g(R)}\ints{2a}{R}{R'} j(R')
=-\frac{{\cal R}_\infty}{g(R)}\ints{2a}{R}{R'} \frac{1}{A_d(R')}\nn\\
&=-\frac{\Gamma(d/2)}{2\pi^{d/2}}\frac{{\cal R}_\infty}{g(R)}\ints{2a}{R}{R'}\frac{1}{R'^{d-1}}\,.
\label{eq: 5.11}
\end{eqnarray}
Thus, a non-vanishing steady-state collision rate is obtained only if $g'(R)\ne f(R)$ or when $d>2$.

\subsection{Collision rate for small particles}
\label{sec: 5.3}

The collision rate must vanish as the particle size approaches zero. This means that the integral in the denominator in (\ref{eq: 5.6}) must diverge for small $a$. Thus if $a\ll\eta$ is small enough, the major contribution in the integral in (\ref{eq: 5.6}) comes from small values of $R$ and the relative error will be small if we replace the integrand by its small $R$ expansion. It is convenient to change variables according to $\bar R=R/\eta$. We expand
\begin{equation}
C(\bar R)=C(0,T)+\frac{1}{2}\partial_{\bar R}^2C(0,T)\bar R^2+\frac{1}{4!}\partial_{\bar R}^4C(0,T)\bar R^{4}+\cdots
\label{eq: 5.12}
\end{equation}
and obtain approximately
\begin{eqnarray}
-\beta^2{\cal N}_d^2f(\bar R)&=-(d+1+(1-d)\Gamma)\DD\eta\bar R+\cdots\nn\\
{\cal N}_d^2g(\bar R)&=-\DD(\eta\bar R)^{2}+\cdots\nn\\
h(\bar R)&=\bar R^{-(d+1+(1-d)\Gamma)}+\cdots\,.
\label{eq: 5.13}
\end{eqnarray}
Here the parameter ${\cal D}$ is defined as
\begin{eqnarray}
\DD&=\frac{1}{2}\left.\frac{{\rm d}^{2}D_{11}}{{\rm d}R^{2}}\right|_{R_1\rightarrow 0,R_{i\ne 1}=0}\nn\\
&=-\frac{d+2}{3!d(1+(d-1)\Gamma)}\frac{u_0^2\eta^{-2}}{\partial_{\bar R}^2C(0,0)}\ints{-\infty}{\infty}{t}\, \partial_{\bar R}^4C(0,t)\,.
\label{eq: 5.14}
\end{eqnarray}
It corresponds to the diffusion constant ${\cal D}_1/(m\gamma)^2$ in equation (39) of \cite{Wil07} and in the incompressible case it corresponds to the diffusion constant ${\cal D}$ defined in equation (17) of \cite{And07}: here $\Gamma$ is an alternative parametrisation of the compressibility, defined as follows
\begin{equation}
\Gamma=\frac{D_{22}(R_1\rightarrow 0,R_{i\ne 1}=0)}{D_{11}(R_1\rightarrow 0,R_{i\ne 1}=0)}
=\frac{d+1+\beta^2}{d-1+3\beta^2}\,.
\label{eq: 4.15}
\end{equation}
The parameter $\Gamma$ was also used in \cite{Wil07} (and earlier work cited therein). It can
take values between
$\Gamma_{\mbox{\tiny{min}}}={1}/{3}$ and
$\Gamma_{\mbox{\tiny{max}}}=({d+1})/({d-1})$.
Here $\Gamma_{\mbox{\tiny{max}}}$ corresponds to a completely
incompressible flow with $\beta=0$ and $\Gamma_{\mbox{\tiny{min}}}$
corresponds to a purely potential ($\beta\rightarrow\infty$) compressible flow. In particular, when $\Gamma=1$ the field strengths of the compressible and incompressible components of the flow are equal, $\beta=1$.

Using equations (\ref{eq: 5.13}), we get the collision rate for small particles
\begin{equation}
{\cal R}_\infty=\frac{2\pi^{d/2}}{\Gamma(d/2)}((d-1)\Gamma-1)\DD n_0\eta^d\lp\frac{2a}{\eta}\rp^{(d-1)\Gamma-1}
\label{eq: 5.16}\,.
\end{equation}
This expression was previously derived in \cite{And07}.

For an incompressible flow, with $\Gamma=(d+1)/(d-1)$, the particle density is uniform and the collision rate is proportional to the packing fraction $n_0a^d$ as expected. By contrast, if the flow is compressible, the particles will cluster on a fractal set \cite{Som93}. This is expected to enhance
the collision rate because the particle density is large within the clusters. The clustering effect can be characterised by the correlation dimension of the particles, $D_2$. It is defined by the scaling law $P(\epsilon)\propto\epsilon^{D_2}$, where $P(\epsilon)$ is the
probability that the particle separation is smaller than $\epsilon$
\cite{Gra83}. We have
\begin{equation}
P(\epsilon)\propto\ints{0}{\epsilon}{R}\,R^{d-1}\,\rho(\ve R)\propto \epsilon^{(d-1)\Gamma-1}
\label{eq: 5.17}
\end{equation}
and thus $D_2=(d-1)\Gamma-1$. This expression
was derived in  \cite{Bal01}.

Equation (\ref{eq: 5.16}) shows that in compressible flows,
the collision rate depends upon the correlation dimension $D_2=(d-1)\Gamma-1<d$. When $\bar a \ll 1$ the collision rate in a compressible flow is therefore much larger than the corresponding rate in an incompressible flow.

Note finally that the steady-state collision rate (\ref{eq: 5.16})
tends to zero when $\Gamma\rightarrow 1/(d-1)$, i.e. when
$\beta \rightarrow\sqrt{d(d-1)/(4-d)}\equiv \beta_{\rm c}$.
For values of $\beta$ larger than $\beta_{\rm c}$,
 no steady-state current satisfying our boundary conditions exists.
In one spatial dimension the parameter $\beta_{\rm c}$ vanishes: no non-trivial steady state exists because all particle trajectories eventually coalesce (this effect was termed \lq path-coalescence' Wilkinson \& Mehlig  \cite{Wil03}).

\subsection{Collision rate for the Gaussian correlation function}
\label{sec: 5.4}

When the correlation function $C(R,T)$ is of the Gaussian form (\ref{eq: 2.6}), we obtain from (\ref{eq: 5.6})
\begin{eqnarray}
{\cal R}_\infty
&=\frac{4\pi^{d/2}n_0\eta^d}{d(d-1+\beta^2)\Gamma(d/2)}\nn\\
&\times\frac{u_0^2\tau}{\eta^2}
\Bigg[\ints{\bar a}{\infty}{\bar R}\bar R^{1-d}e^{\bar R^2/2}\Bigg\{\bigg[(d-1+\beta^2)\lp e^{\bar R^2/2}-1\rp+\beta^2\bar R^2\bigg]\nn\\
&\times\exp\lp\beta^2\ints{\bar R}{\infty}{R'} R' \frac{2+d-R'^2}{(d-1+\beta^2)\lp e^{R'^2/2}-1\rp+\beta^2R'^2}\rp\Bigg\}^{-1}
\Bigg]^{-1}
\label{eq: 5.18}
\end{eqnarray}
where as before $R=\eta\bar R$. The major contribution to the integral comes from small values of $\bar R$ (except for when $d=2$ and $\beta=1$ or $d=1$, when the integral diverges) and for small values of $\bar a$, the integrand can be expanded in powers of $\bar R$. Expanding the exponential function in the integrand yields
\begin{equation}
{\cal R}_\infty=\frac{2((d-1)\Gamma-1)\pi^{d/2}}{\Gamma(d/2)}\DD n_0\eta^d\bar a^dh(\bar a)
\label{eq: 5.19}
\end{equation}
where the function $h$ was defined in (\ref{eq: 5.3}).
The diffusion constant ${\cal D}$ is given by (\ref{eq: 5.14}). With  (\ref{eq: 2.6}) we find
\begin{equation}
\DD=\frac{d-1+3\beta^2}{d(d-1+\beta^2)}\frac{u_0^2\tau}{\eta^2}\,.
\label{eq: 5.20}
\end{equation}
For low dimensions,
\begin{equation}
h(\bar a)\approx \bar a^{(d-1)\Gamma-d-1}
\label{eq: 5.21}
\end{equation}
is a fairly good approximation (a maximum of 3 percent error when $d=2$).
Substituting (\ref{eq: 5.21})  into (\ref{eq: 5.19}) gives once more (\ref{eq: 5.16}).

\section{Time-dependent solution}
\label{sec: 6}

So far we have derived the probability densities and collision rates for the steady state. We now wish to derive expressions of these quantities as functions of time. To this end we need to solve the time dependent Fokker-Planck equation (\ref{eq: 4.9}) with the boundary conditions
\begin{equation}
\rho(2a,t)=0\,,\quad\mbox{and}\quad
\rho(\infty,t)=n_0\,.
\label{eq: 6.1}
\end{equation}
As before we assume a uniform initial scatter of particles
\begin{equation}
\rho(R,0)=\Theta(R-2a)\,n_0\,.
\label{eq: 6.2}
\end{equation}
The collision rate is given by equation (\ref{eq: 4.8})
\begin{equation}
{\cal R}(t)
=-\frac{2\pi^{d/2}(2a)^{d-1}}{\Gamma(d/2)}j_r(R=2a,t)\,,
\label{eq: 6.3}
\end{equation}
where we have used that the area of a $d$-dimensional sphere of radius $r$ is $A_d(r)=2\pi^{d/2}r^{d-1}/\Gamma(d/2)$, and the gamma function is denoted by $\Gamma(z)$.

In order to render the boundary conditions homogeneous, we split
$\rho(R,t)=\rho_1(R,t)+\rho_2(R)$ and impose the conditions
$\rho_1(2a,t)=\rho_1(\infty,t)=0$, and that the differential equation homogeneous in $\rho$ is also homogeneous in $\rho_1$. Thus $\rho_2(R)$ is uniquely given by the steady-state solutions found in the previous section.

Now consider the remaining equation for $\rho_1(R,t)$
which is identical to (\ref{eq: 4.9}) with homogeneous boundary conditions and initial condition
\begin{equation}
\rho_1(R,0)=\Theta(R-2a)(n_0-\rho(R,\infty))\,.
\label{eq: 6.4}
\end{equation}
Separation of variables $\rho_1(R,t)=\mathcal{F}(R)\mathcal{G}(t)$ in (\ref{eq: 4.9})
gives (using the radial current (\ref{eq: 4.10}))
\begin{equation}
\frac{1}{\mathcal{G}}\frac{\partial\mathcal{G}}{\partial t}= \frac{1}{R^{d-1}\mathcal{F}}\frac{\partial}{\partial R}\lp R^{d-1}\left[f(R)\mathcal{F}+g(R)\frac{\partial}{\partial R}\mathcal{F}\right]\rp\,.
\label{eq: 6.5}
\end{equation}
Since the left-hand side depends on $t$ only, and the right-hand side
depends on $R$ only, both sides must be equal to a constant, $B$ say.
Choosing $B=-\DD\mu^2$, where $\mu$ is a positive dimensionless constant, we obtain
\begin{equation}
\mathcal{G}(t)=\mathcal{G}_0e^{-\mu^2\DD t}\,.
\label{eq: 6.6}
\end{equation}
Note that the solution corresponding to $\mu=0$ has already been taken care of in $\rho_2$.
Therefore  we can restrict ourselves to considering $\mu>0$ in the following.

To solve the radial part of (\ref{eq: 6.5}), we consider the limit of $R\ll\eta$, where $f(R)$ and $g(R)$ are simple power laws
(this follows from (\ref{eq: 5.13})).
Using the dimensionless variable $\bar R=R/\eta$ gives the following approximate equation
for  small values of $\bar R$
\begin{equation}
\bar R^2\frac{\partial^2\mathcal{F}}{\partial\bar R^2} + \lp 2d+2+(1-d)\Gamma\rp\bar R\frac{\partial\mathcal{F}}{\partial\bar R} + (d+1+(1-d)\Gamma)d\mathcal{F} = -\mu^2\mathcal{F}.
\label{eq: 6.7}
\end{equation}
This is an Euler equation. Its solution is obtained by the variable substitution $\bar R=\bar ae^u$,
where the factor $\bar a=2a/\eta$ is included in for later convenience and $0\le u<\infty$. We find
\begin{equation}
\mathcal{F}(\bar R)=\bar R^{-d-\mu_0}\cdot\lbp\begin{array}{ll}
D_-\sin\lp\sqrt{\mu^2-\mu_0^2}\ln(\bar R)+\phi_-\rp,\mbox{ if }\mu^2\ge\mu_0^2\cr
D_+\sinh\lp\sqrt{\mu_0^2-\mu^2}\ln(\bar R)+\phi_+\rp,\mbox{ if }0<\mu^2<\mu_0^2
\end{array}\right.
\label{eq: 6.8}
\end{equation}
where
\begin{equation}
\mu_0=\frac{1}{2}(1+(1-d)\Gamma)  = \frac{d(1-d)+(4-d)\beta^2}{2(d-1+3\beta^2)}
\label{eq: 6.9}
\end{equation}
and $D_\pm$ and $\phi_\pm$ are integration constants.

For all values of $\mu_0$ in the allowed range $-\frac{d}{2}\le\mu_0\le\frac{1}{6}(4-d)$, the boundary condition $\mathcal{F}(\infty)=0$ is automatically fulfilled. The boundary condition
$\mathcal{F}(\bar a)=0$ gives $\phi_\pm=-\sqrt{\pm(\mu_0^2-\mu^2)}\ln(\bar a)+\sqrt{\mp 1}n\pi$, and thus
\begin{equation}
\mathcal{F}(\bar R)=\bar R^{-d-\mu_0}\cdot\lbp\begin{array}{ll}
D_-\sin\lp\sqrt{\mu^2-\mu_0^2}\ln(R/\bar a)\rp,\mbox{ if }\mu^2\ge\mu_0^2\cr
D_+\sinh\lp\sqrt{\mu_0^2-\mu^2}\ln(R/\bar a)\rp,\mbox{ if }0<\mu^2<\mu_0^2
\end{array}\right..
\label{eq: 6.10}
\end{equation}
The boundary conditions do not constrain the eigenvalue $\mu$
and we need to consider a continuous superposition of eigenfunctions
\begin{eqnarray}
\rho_1(\bar R,t)&=\bar R^{-d-\mu_0}\ints{|\mu_0|}{\infty}{\mu}\,\zeta(\mu)\sin\lp\sqrt{\mu^2-\mu_0^2}\ln(R/\bar a)\rp e^{-\mu^2\DD t}
\nn\\&+\bar R^{-d-\mu_0}\ints{0^+}{|\mu_0|}{\mu}\, \tilde\zeta(\mu)\sinh\lp\sqrt{\mu_0^2-\mu^2}\ln(R/\bar a)\rp e^{-\mu^2\DD t}\,.
\label{eq: 6.11}
\end{eqnarray}
Here $\zeta(\mu)$ and $\tilde\zeta(\mu)$ are functions to be determined by the initial conditions.
It turns out that it is sufficient
to consider $\zeta(\mu)$ for our initial condition and
take $\tilde\zeta(\mu)=0$ from now on.

At time $t=0$ we make the change of variables $u=\ln(\bar R/\bar a)$ as before. We set $s=\sqrt{\mu^2-\mu_0^2}$  and find
\begin{equation}
\rho_1(u,0)=\frac{{\rm d}R}{{\rm d}u}(\bar ae^u)^{-d-\mu_0}\ints{0}{\infty}{s}\,\frac{{\rm d}\mu}{{\rm d}s}\zeta(s)\sin\lp su\rp.
\label{eq: 6.12}
\end{equation}
Comparing this to the required initial distribution
\begin{equation}
\rho_1(u,0)=\frac{{\rm d}R}{{\rm d}u}\Theta(u)(n_0-\rho(u,\infty))
\label{eq: 6.13}
\end{equation}
gives
\begin{equation}
\label{eq: 6.14}
\ints{0}{\infty}{s}\,\frac{{\rm d}\mu}{{\rm d}s}\zeta(s)\sin\lp su\rp=(\bar ae^u)^{d+\mu_0}\Theta(u)(n_0-\rho(u,\infty)).
\end{equation}
The left hand side is of the form of a Fourier sine transform with inverse
\begin{equation}
\label{eq: 6.15}
\frac{{\rm d}\mu}{{\rm d}s}\zeta(s)=\frac{2}{\pi}\bar a^{d+\mu_0}\ints{0}{\infty}{\tilde u}\, (n_0-\rho(\tilde u,\infty))e^{(d+\mu_0)\tilde u}\sin(s\tilde u).
\end{equation}
We insert this expression into the ansatz for $\rho_1$. Upon
changing  order of the integrations
(that is we perform the $s$-integral first), we find
\begin{eqnarray}
\rho_1(u,t)
&=\frac{{\rm d}R}{{\rm d}u}\frac{1}{\sqrt{\pi\DD t}}e^{-\mu_0^2\DD t-(d+\mu_0)u-u^2/(4\DD t)}\nn\\
&\hspace*{1cm}\times\ints{0}{\infty}{\tilde u}\, (n_0-\rho(\tilde u,\infty))e^{(d+\mu_0)\tilde u -\tilde u^2/(4\DD t)}\sinh\lp\frac{u\tilde u}{2\DD t}\rp\,.
\label{eq: 6.16}
\end{eqnarray}
To calculate the collision rate we
need to know the current $j_r$ at radius $2a$. To this end we
just require the derivative of $\rho$ evaluated at $R=2a$ (since $\rho$ itself is constrained to vanish there).
We find
\begin{equation}
\frac{\partial\rho_1}{\partial R}(2a,t)=
\frac{e^{-\mu_0^2\DD t}}{4a\sqrt{\pi\DD^3t^3}}\ints{0}{\infty}{\tilde u}\, (n_0-\rho(\tilde u,\infty))\,\tilde u\,e^{(d+\mu_0)\tilde u - \tilde u^2/(4\DD t)}
\label{eq: 6.17}
\end{equation}
Equation (\ref{eq: 6.3}) now allows us to calculate the time-dependent collision rate. We split $j_r$ [see equation (\ref{eq: 4.10})]
into two parts depending on $\rho_1$ and $\rho_2$ respectively. The second part gives the steady-state collision rate ${\cal R}_\infty$, see equation (\ref{eq: 5.6}). We find (expanding in powers of $\bar R$)
\begin{eqnarray}
{\cal R}(t)
&={\cal R}_\infty+\frac{\pi^{(d-1)/2}(2a)^{d}}{\Gamma(d/2)}\frac{e^{-\mu_0^2\DD t}}{\sqrt{\DD t^3}}\nn\\
&\hspace*{1cm}\times\ints{0}{\infty}{\tilde u}\, (n_0-\rho(\tilde u,\infty))\,\tilde u\,e^{(d+\mu_0)\tilde u - \tilde u^2/(4\DD t)}\,.
\label{eq: 6.18}
\end{eqnarray}
This is our main result for the time-dependent collision rate,
valid for correlations with non-vanishing fourth order derivative at $R=0$ and for small particles.

An approximation to this result can be obtained by approximating the steady-state probability density
\begin{equation}
\rho(\bar R,\infty)=n_0\eta h(\bar R)\ints{\bar a}{\bar R}{\bar R'}\frac{\bar R'^{1-d}}{g(\bar R')h(\bar R')}\Bigg/\ints{\bar a}{\infty}{\bar R'}\frac{\bar R'^{1-d}}{g(\bar R')h(\bar R')}
\label{eq: 6.19}
\end{equation}
by expanding in powers of $\bar R$.
Since the $\tilde u$-integral extends to infinity and since the Gaussian contribution in the integrand becomes small for large values of $\DD t$, we must have that $\rho(\tilde u,\infty)$ goes to $n_0$ as $\tilde u$ goes to $\infty$, which is the case for the original $\rho(\tilde u,\infty)$.
To accomplish this, we match the small-$\bar R$ expression of $\rho(\tilde u,\infty)$ at $\bar R=1$ to the large-$\bar R$ expression.
By using (\ref{eq: 5.13}) and (\ref{eq: 5.16}) and matching at $\bar R=1$ we find
\begin{equation}
\label{eq: 6.20}
\rho(\bar R,\infty)=n_0\eta\lp \Theta(\bar R-1)+\bar R^{-d-2\mu_0}(1-\Theta(\bar R-1))-\bar a^{-2\mu_0}\bar R^{-d}\rp\,,
\end{equation}
where the last term is not cut off at $\bar R=1$ as it vanishes sufficiently quickly.

The time-dependent collision rate can now be evaluated from the sum of three integrals of the type
\begin{eqnarray}
\label{eq: 6.21}
I_n&=a_n\ints{0}{c_n}{\tilde u}\, \tilde u\,e^{(b_n+d+\mu_0)\tilde u - \tilde u^2/(4\DD t)}\nn\\
&=2\DD t a_n\bigg[
1-e^{(b_n+d+\mu_0)c_n-c_n^2/(4\DD t)}+\sqrt{\pi\DD t}(b_n+d+\mu_0)e^{(b_n+d+\mu_0)^2\DD t}\nn\\
&\hspace*{1mm}\times\lp\mathrm{erf}((b_n+d+\mu_0)\sqrt{\DD t})
-\mathrm{erf}\Big((b_n+d+\mu_0)\sqrt{\DD t}-\frac{c_n}{2\sqrt{\DD t}}\Big)\rp
\bigg]
\end{eqnarray}
where $a_n=\{1,-\bar a^{-d-2\mu_0},\bar a^{-d-2\mu_0}\}$, $b_n=\{0,-d-2\mu_0,-d\}$ and $c_n=\{-\ln(\bar a),-\ln(\bar a),\infty\}$. Putting everything together we find
\begin{eqnarray}
\label{eq: 6.22}
{\cal R}(t)
&=2\frac{\pi^{d/2}}{\Gamma(d/2)}n_0\eta^d\DD
\Bigg\{
-\mu_0\bar a^{-2\mu_0}\lp 1+\mathrm{erf}\left[-\mu_0\sqrt{\DD t}+\frac{\ln\bar a}{2\sqrt{\DD t}}\right]\rp\nn\\
&+\bar a^{d}\Bigg( \frac{1}{\sqrt{\pi\DD t}}e^{-\mu_0^2\DD t}+(d+\mu_0)e^{d(d+2\mu_0)\DD t}\nn\\
&\hspace*{1mm}\times\lbp\mathrm{erf}\left[(d+\mu_0)\sqrt{\DD t}\right]-\mathrm{erf}\left[(d+\mu_0)\sqrt{\DD t}+\frac{\ln\bar a}{2\sqrt{\DD t}}\right]\rbp\Bigg)
\Bigg\}
\end{eqnarray}
valid for $\bar a\ll 1$.  For times less than $[d(d+2\mu_0)\DD]^{-1}$
the above expression simplifies to
\begin{eqnarray}
\label{eq: 6.23}
{\cal R}(t)&\approx&-\frac{1}{2\mu_0}{\cal R}_\infty\bar a^{d+2\mu_0}\frac{1}{\sqrt{\pi\DD t}}e^{-\mu_0^2\DD t}\,.
\end{eqnarray}
For intermediate times, $[d(d+2\mu_0)\DD]^{-1} \ll t \ll \ln(\bar a)/(2\mu_0\DD)$,
we find approximately
\begin{eqnarray}
\label{eq: 6.24}
 {\cal R}(t)&\approx&-\frac{1}{2\mu_0}{\cal R}_\infty\bar a^{d+2\mu_0}(d+\mu_0)e^{d(d+2\mu_0)\DD t}\nn\\
 &&\hspace*{1mm}\times\lbp\mathrm{erf}\left[(d+\mu_0)\sqrt{\DD t}\right]-\mathrm{erf}\left[(d+\mu_0)\sqrt{\DD t}+\frac{\ln\bar a}{2\sqrt{\DD t}}\right]\rbp\Bigg)\,,
\end{eqnarray}
while for large times, $t \gg {\rm ln}(\bar a)/(2\mu_0\DD)$, where
\begin{eqnarray}
\label{eq: 6.25}
{\cal R}(t)&\approx&\frac{1}{2}{\cal R}_\infty\lp 1+\mathrm{erf}\left[-\mu_0\sqrt{\DD t}+\frac{\ln\bar a}{2\sqrt{\DD t}}\right]\rp\,.
\end{eqnarray}
Consider finally evaluating (\ref{eq: 6.22}) in the
incompressible limit, we have $\mu_0=-d/2$ and the terms involving the integral cutoffs $\ln(\bar a)$ cancel ($I_1=-I_2$ in (\ref{eq: 6.21})).
Thus in the case $\beta=0$ the collision rate simplifies to
\begin{equation}
\label{eq: 6.26}
{\cal R}(t)=d\frac{\pi^{d/2}(2a)^{d}}{\Gamma(d/2)}n_0\DD\Bigg\{1+\frac{2}{d\sqrt{\pi\DD t}}e^{-\frac{1}{4}d^2\DD t}+\mathrm{erf}\left[\frac{d}{2}\sqrt{\DD t}\right]\Bigg\}.
\end{equation}
Equations (\ref{eq: 6.22}) and (\ref{eq: 6.26}) are compared to results of numerical simulations in the following section.


\section{Numerical illustration and concluding remarks}
\label{sec: 7}

\begin{figure}[t]
\begin{center}
\includegraphics[width=12cm]{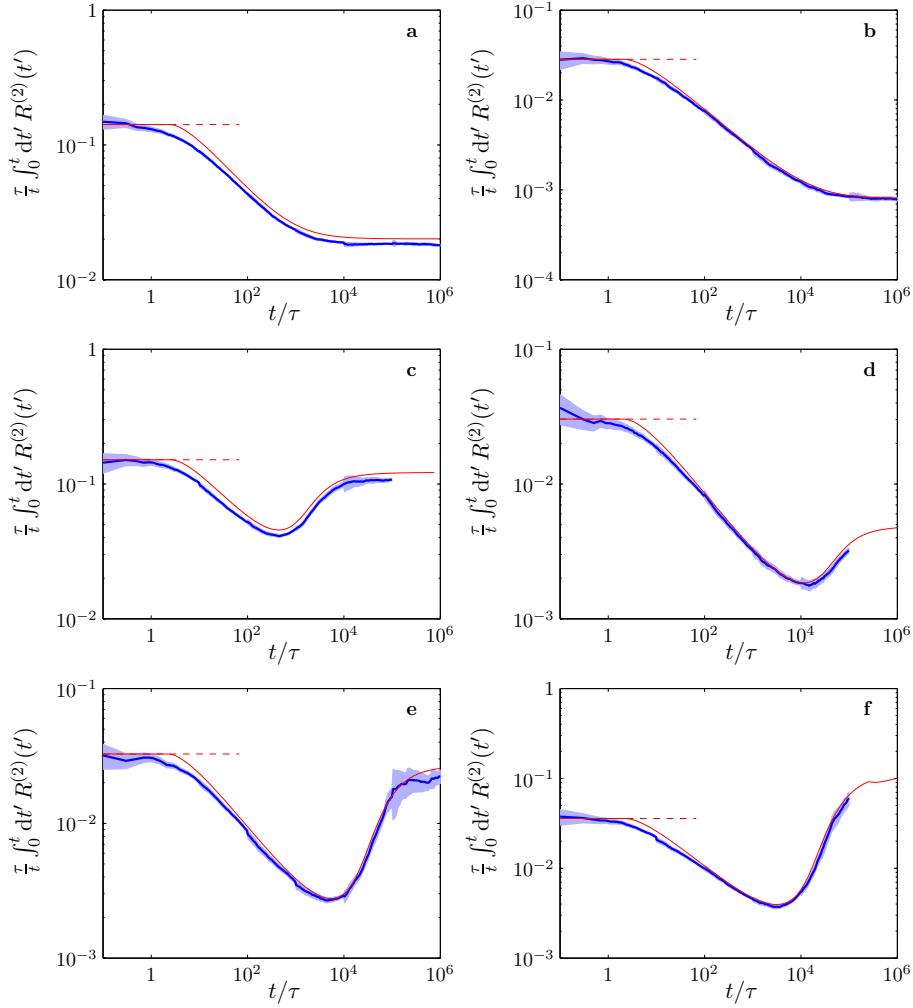}
\caption{\label{fig: 3}
Collision rate for particles advected in a two-dimensional
flow of the form  (\ref{eq: 2.2})
with (\ref{eq: 2.6}) and  $C_0=u_0^2\eta^2/2$.
Particles are initially randomly distributed, initially overlapping particle pairs are removed, as are particles which have collided.
Parameters: $n_0=1000$, $u_0=1$, $\eta=0.1$ and $a=0.001$ for all graphs and for {\bf a} $\beta=0$ and $\tau=0.004$, {\bf b} $\beta=0$ and $\tau=0.0008$, {\bf c} $\beta=1/\sqrt{13}$ and $\tau=0.004$, {\bf d} $\beta=1/\sqrt{13}$ and $\tau=0.0008$, {\bf e} $\beta=1/\sqrt{5}$ and $\tau=0.0008$, {\bf f} $\beta=\sqrt{3/7}$ and $\tau=0.0008$.
The collision rate is approximated by the cumulative sum of all collisions up to $t$,
 divided by $t$.
The light blue areas are the intervals $R\pm 2\sigma/\sqrt{N}$,
where $\sigma$ is the standard deviation of the rate, and $N$ is the number of realisations
of the flow. The latter decreases over time (it is typically $N\sim 10^4$ for the
first two decades in $t$, and $N\sim 10$ for the last decade).
The Saffman-Turner estimates (\ref{eq: 3.8},\ref{eq: 3.9}) are shown as red dashed lines.
Our own result (\ref{eq: 6.22}) is shown as red solid lines.
To correspond to the simulated collision rate, the plotted theoretical collision rates has been integrated to and then divided by $t$.
Due to the assumptions in the Fokker-Planck theory, the long time rate (\ref{eq: 6.22}) is not valid for $t<\tau$. The plotted long time theory is a combination of the short time rate (\ref{eq: 7.1}) and the long time rate (\ref{eq: 6.22}), matched at the time at which the long time rate drops below the short time rate (which is an upper bound).
}
\end{center}
\end{figure}

We performed simulations of the collision rate as a function
of time at small Kubo numbers, for both incompressible and compressible flows. These illustrate (figure \ref{fig: 3}) the very complex behaviour of the model. For example,  in the case of compressible flows the collision rate at first decreases below the value determined by the Saffman-Turner approximation due to the effect illustrated in figure \ref{fig: 2}, before rising again as particle clustering becomes apparent.

\subsection{Simulations}
\label{sec: 7.1}

Figure \ref{fig: 3} summarises our results for the collision rate of particles advected
in flows with small Kubo numbers.  Shown are numerical simulations
for a two-dimensional random flow of the form (\ref{eq: 2.2})
with correlation function (\ref{eq: 2.6}).

Panels {\bf a} and {\bf b} show
the collision rate in an incompressible flow ($\beta=0$).
As expected (see section \ref{sec: 3.2}), the collision rates drops below
the initial transient given by the Saffman-Turner approximation,
(\ref{eq: 3.8},\ref{eq: 3.9})
with $\beta = 0$. For the particular choice (\ref{eq: 2.6}) with
$C_0=u_0^2\eta^2/2$, equations (\ref{eq: 3.8},\ref{eq: 3.9})
become in two spatial dimensions
\begin{equation}
\label{eq: 7.1}
{\cal R}_0= \sqrt{\pi}(2a)^2n_0\frac{u_0}{\eta}\sqrt{\frac{1+3\beta^2}{1+\beta^2}}\,.
\end{equation}
Also shown is our own theory valid for small Kubo numbers (equation (\ref{eq: 6.22}) in section \ref{sec: 6}).  The agreement between the theory and the simulations is good in all cases, but slightly better for the smaller Kubo number (panel {\bf b}). As discussed in section \ref{sec: 3.2}, the initial transient  constitutes an upper bound to the collision rate.

Panels {\bf c}  to {\bf f}
in figure \ref{fig: 3} show the collision rate in compressible flow.
Now, the Saffman \& Turner theory is no longer an upper bound
(see for example figure \ref{fig: 3}{\bf f}),
because initially homogeneously distributed particles in an incompressible
flow cluster together. The corresponding density fluctuations increase
the collision rate, as our exact result shows (equation (\ref{eq: 6.22}) in section \ref{sec: 6}).
Again we observe good agreement between the simulations and our analytical
result.

\subsection{Scope for further investigations}
\label{sec: 7.2}

In this paper we have concentrated upon the solvable case of advective collisions in flows with small Kubo number, which provides considerable physical insight. We conclude by commenting on the relation between these results and collisions in a turbulent flow field which satisfies the Navier-Stokes equation. The standard approach, based upon the Saffman-Turner formula, predicts a collision rate ${\cal R}\sim n_0a^d \tau^{-1}$, where $\tau$ is the Kolmogorov timescale \cite{Fri97} of the turbulent flow. The calculation based upon the diffusion equation gives a collision rate ${\cal R}\sim {\rm Ku}\,n_0a^d\tau^{-1}$. Given that ${\rm Ku}=O(1)$ for a turbulent flow, we see that the Saffman-Turner and diffusive expressions are of the same order.

These observations are consistent with the hypthesis that
the collision rate for small particles in a turbulent flow is
\begin{equation}
\label{eq: 7.2}
{\cal R}=K_d \frac{n_0a^d {\cal E}^{1/2}}{\nu^{1/2}}\,,
\end{equation}
where ${\cal E}$ is the rate of dissipation per unit mass, $\nu $ is the kinematic viscosity, and $K_d$ is a universal constant (depending only on the dimension). It would be a valuable addition to the literature on aerosols and suspended
 particles to determine the value of $K_3$ from numerical simulations using a Navier-Stokes flow.

{\em Acknowledgments.} We acknowledge support from
Vetenskapsr\aa{}det and from  the research initiative
\lq Nanoparticles in an interactive
environment' at G\"oteborg university.
\mbox{}\\[1cm]


\newpage
\appendix


\section{Evaluation of collision rate for constant shear}
\label{sec: appA}

In this appendix we show how to evaluate the expression (\ref{eq: 3.2})
for a general matrix $\ma A$ with elements $A_{ij}$ (we drop the subscript $0$). It is convenient to decompose $\ma A$ into  a symmetric part
$\ma{A}_{()} $ and an antisymmetric part $\ma{A}_{[]}$
\begin{equation}
\label{eq: A.1}
\ma{A}
= \lp
\begin{array}{ccc}
A_{11} & \frac{A_{12}+A_{21}}{2} & \\
\frac{A_{12}+A_{21}}{2} & A_{22} &  \\
\vdots &  & \ddots \\
\end{array}
\rp
+
\lp
\begin{array}{ccc}
0 & \frac{A_{12}-A_{21}}{2} & \\
-\frac{A_{12}-A_{21}}{2} & 0 & \\
\vdots &  & \ddots \\
\end{array}\rp \,.
\end{equation}
The collision rate is independent of the antisymmetric
part $\ma{A}_{[]}$ (because rotations do not contribute to the collision rate). The symmetric part $\ma{A}_{()}$ can be diagonalised by an
orthogonal transformation, $\ma A_{()}=\ma{O}\ma{S}\ma{O}^T$, where
$\ma S$ is diagonal with the eigenvalues
$\sigma_1\le\sigma_2\le...\le\sigma_n$ of $\ma A_{()}$.
In evaluating (\ref{eq: 3.2}) we may write
$\hat{\ve n}^T\ma A\hat{\ve n}
= \hat{\ve n}^T\ma O\ma S\ma O^T\hat{\ve n} = \hat{\ve n}'^T\ma S\hat{\ve n}'$, where $\ve{\hat n}'=\ma O^T\ve{\hat n}$.
Since $\ve{\hat n}'$ is just a rotation of $\ve{\hat n}$ which is integrated over all directions, we can replace (\ref{eq: 3.2}) by
\begin{equation}
\label{eq: A.2}
{\cal R}_0
= -2an_0\ints{}{}{\Omega}\ \ve{\hat n}^T\ma{S}\ve{\hat n}\ \Theta(-\ve{\hat n}\ma{S}\ve{\hat n})\,.
\label{eq:nSn}
\end{equation}
%
%


\subsection{Two spatial dimensions}
\label{sec: appA.1}

We now show how to perform the integral in (\ref{eq:nSn}) in two spatial dimensions.
Note that if the flow determined by $\ma A$ is not area preserving,
the particle density will change as a function of time. In this case
$n_0$ in (\ref{eq: 3.2}) must be replaced by
\begin{equation}
\label{eq: A.3}
n(t) = n_0\frac{A_0}{A(t)} = n_0\exp[-\tr\ma{A}t]\,.
\end{equation}
At short times we approximate $n(t) \approx n_0$ and find
\begin{eqnarray}
\label{eq: A.4}
{\cal R}(\sigma_+,\sigma_-) \nn\\
&&\hspace*{-2.2cm}\approx -(2a)^2n_0\ints{0}{2\pi}{\theta}\, (\sigma_1\cos^2{\theta}+\sigma_2\sin^2\theta)\,
\Theta(-\sigma_1\cos^2{\theta}-\sigma_2\sin^2\theta) \nn\\
&&\hspace*{-2.2cm}= 2(2a)^2n_0
\left\{
\begin{array}{ll}
-\pi\sigma_+ & \mbox{if }\sigma_+<-\sigma_-\\
\sqrt{\sigma_-^2-\sigma_+^2}-\sigma_+\arccos({\sigma_+}/{\sigma_-}) & \mbox{if }-\sigma_-<\sigma_+<\sigma_-\\
0 & \mbox{if }\sigma_+>\sigma_-\\
\end{array}
\right.
\end{eqnarray}
where  $\sigma_+=(\sigma_1+\sigma_2)/2$ and $\sigma_-=(\sigma_2-\sigma_1)/2>0$. This is the final result in two spatial dimensions, expressed in terms of the eigenvalues $\sigma_1 \leq \sigma_2$ of the symmetric part $\ma A_{()}$ of the strain matrix.


\subsection{Three spatial dimensions}

In three spatial dimensions we only consider incompressible shear flows,
that is a general, three-dimensional traceless matrix $\ma{A}_{()}$.
It has eigenvalues $\sigma_3\ge\sigma_2\ge\sigma_1$,
obeying the relations $\sigma_3\ge 0$, $-{\sigma_3}/{2}\le\sigma_2\le\sigma_3$ and $\sigma_1=-\sigma_2-\sigma_3$.
We use a spherical coordinate system,
$\hat{\ve n}=(\sin\theta\cos\varphi,\sin\theta\sin\varphi,\cos\theta)$ and find
\begin{eqnarray}
\label{eq: A.5}
\hat{\ve n}^T\ma S \hat{\ve n} &=& (-\sigma_2-\sigma_3)\sin^2\theta\cos^2\varphi+\sigma_2\sin^2\theta\sin^2\varphi+\sigma_3\cos^2\theta\nn\\
& = & \sigma_3\lbp\frac{3}{2}\cos^2\theta-\frac{1}{2} -\lp\frac{\sigma_2}{\sigma_3}+\frac{1}{2} \rp\lp 1-\cos^2\theta \rp\cos(2\varphi) \rbp\,.
\end{eqnarray}
Substitute $t=\cos\theta$ and $2\varphi\rightarrow\varphi'=\varphi$ we obtain
\begin{eqnarray}
\label{eq: A.6}
{\cal R}_0=(\sigma_2,\sigma_3) &=&-\ints{0}{2\pi}{\varphi}\ints{0}{\pi}{\theta} \sin(\theta)\hat{\ve n}^T \ma S\hat{\ve n} \Theta(-\hat{\ve n}^T \ma S\hat{\ve n}) \nn\\
 &=&-2\ints{0}{2\pi}{\varphi}\ints{0}{1}{t} \hat{\ve n}^T \ma S\hat{\ve n} \Theta(-\hat{\ve n}^T \ma S\hat{\ve n}),
\end{eqnarray}
where the transformed integrand
\begin{equation}
\label{eq: A.7}
\hat{\ve n}^T \ma S\hat{\ve n} = \sigma_3\lbp\frac{3}{2}t^2-\frac{1}{2} -\lp\frac{\sigma_2}{\sigma_3}+\frac{1}{2} \rp\lp 1-t^2 \rp\cos\varphi\rbp
\end{equation}
is smaller than $0$ for $t<t_0$, where $t_0$ is given by
\begin{equation}
\label{eq: A.8}
t_0^2 = \frac{1+x\cos\varphi}{3+x\cos\varphi}\,,
\end{equation}
and $x=\lp 2{\sigma_2}/{\sigma_3}+1\rp$. Because the values $\sigma_2$ can take are limited by $\sigma_3$, $x$ must be in the range $0\le x\le 3$ and $t_0$ in the range $-\infty\le t_0^2\le \frac{1}{2}$. Performing the $t$-integral from $0$ to $t_0$ gives
\begin{eqnarray}
\label{eq: A.9}
{\cal R}_0(\sigma_2,\sigma_3) & =& \ints{0}{2\pi}{\varphi} \sigma_3
\left\{
\begin{array}{ll}
-t_0^3+t_0+x\left(t_0-\frac{t_0^3}{3}\right) \cos\varphi & \mbox{if $t_0>0$}\\
0 & \mbox{if $t_0<0$}\\
\end{array}
\right.  \nn\\
& = &\frac{2}{3}\sigma_3\ints{-\varphi_0}{\varphi_0}{\varphi} \frac{\left(1+x\cos\varphi\right)^{3/2}}{\left(3+x\cos\varphi\right)^{1/2}}
\eqnlab{collrate_3d_integrand_phi}
\end{eqnarray}
where $\varphi_0=\arccos(-1/x)$, if $\sigma_2>0$ and $\varphi_0=\pi$, if $\sigma_2\le 0$.

Expanding the integrand around in $x$ around the point $x_0$ (to be determined below) we find
\begin{eqnarray}
\label{eq: A.10}
I_2  &=& 2\sigma_3\sum_{n=0}^\infty\sum_{k=0}^{n}\left(\frac{n}{k}\right)\cos^{n}\varphi (-2)^{-n}(2n-2k-5)!!(2k-1)!!\nn\\
&&\hspace{-2mm}\times\left(1+x_0\cos\varphi\right)^{3/2-n+k}\left(3+x_0\cos\varphi\right)^{-1/2-k}\frac{(x-x_0)^n}{n!}\,.
\end{eqnarray}
If we choose $x_0=0$, which corresponds to $\sigma_2=-\sigma_3/2<0$, we can perform the integration after changing the
order of summation and integration. The collision rate becomes
\begin{eqnarray}
\label{eq: A.11}
{\cal R}_0(\sigma_2<0,\sigma_3)\\
&&\hspace*{-2cm}  = \frac{4\pi}{\sqrt{3}}\sigma_3\sum_{n=0}^\infty\sum_{k=0}^{2n}\frac{(4n-2k-5)!!(2k-1)!!(2n-1)!!}{2^n3^{k}(2n-k)!k!n!}\lp \frac{\sigma_2}{\sigma_3}+\frac{1}{2}\rp^{2n}\nn\,.
\end{eqnarray}
Only even powers contribute to this sum:
we have not used that $\sigma_2\ge\sigma_1$ and could thus as well have
expanded the starting equations using $\sigma_1$ and $\sigma_3$,
with the only difference that $\sigma_2$ would be replaced by $-\sigma_3-\sigma_1$ in (\ref{eq: A.11}). We thus find that ${\cal R}_0(\sigma_2,\sigma_3)$ is symmetric around $\sigma_1=\sigma_2=-\sigma_3/2$, i.e. ${\cal R}_0(\sigma_2,\sigma_3)={\cal R}_0(-\sigma_2-\sigma_3,\sigma_3)$.

To obtain an expression valid for $\sigma_2>0$, we reorder the eigenvalues of the matrix $\ma{A}_{()}$ as $\sigma_3\rightarrow\sigma_1$ and $\sigma_1\rightarrow\sigma_3=-\sigma_2-\sigma_1$, giving the eigenvalue ranges $\sigma_1\le 0$ and $\sigma_1\le\sigma_2\le-\sigma_1/2$.
The integrand analogous to (\ref{eq: A.5}) becomes
\begin{equation}
\label{eq: A.12}
\hat{\ve n}^T \ma S\hat{\ve n} = \sigma_1\lbp\frac{3}{2}t^2-\frac{1}{2} -\lp\frac{\sigma_2}{\sigma_1}+\frac{1}{2} \rp\lp 1-t^2 \rp\cos\varphi\rbp,
\end{equation}
where $\sigma_1$ has opposite sign as $\sigma_3$ before.
This integrand is smaller than $0$ for $t>t_0$, where $t_0$ is given by
\begin{equation}
\label{eq: A.13}
t_0^2 = \frac{1+y\cos\varphi}{3+y\cos\varphi}
\end{equation}
where $y=\lp 2{\sigma_2}/{\sigma_1}+1\rp$ lies in the interval $0\le y\le 3$ and $t_0$ lies in $-\infty\le t_0^2\le {1}/{2}$. Performing the $t$ integral from $t_0$ to $1$ gives
\begin{eqnarray}
\label{eq: A.14}
{\cal R}_0(\sigma_2,\sigma_3) & =& \ints{0}{2\pi}{\varphi} \sigma_1
\left\{
\begin{array}{ll}
\frac{2}{3}y\cos\varphi+t_0^3-t_0+y\left(\frac{t_0^3}{3}-t_0\right) \cos\varphi & \mbox{if $t_0>0$}\\
\frac{2}{3}y\cos\varphi & \mbox{if $t_0<0$}\\
\end{array} \right.  \nn\\
&=& -\frac{2}{3}\sigma_1\ints{-\varphi_0}{\varphi_0}{\varphi} \frac{\left(1+y\cos\varphi\right)^{3/2}}{\left(3+y\cos\varphi\right)^{1/2}}
\end{eqnarray}
where we have proceeded as in deriving (\ref{eq: A.12}) with $\varphi_0=\arccos(-1/y)$, if $\sigma_2<0$ and $\varphi_0=\pi$, if $\sigma_2\ge 0$. We obtain the final result
\begin{eqnarray}
\label{eq: A.15}
{\cal R}_0(\sigma_2,\sigma_3)&=&n_0 (2a)^3 \!\left\{
\begin{array}{l}
\frac{4\pi\sigma_3}{\sqrt{3}}{\displaystyle \sum_{n=0}^\infty} c_n \lp\frac{1}{2}+\frac{\sigma_2}{\sigma_3}\rp^{2n} \,\,  \mbox{if $\sigma_2 \leq 0$}\\
\frac{4\pi(\sigma_2+\sigma_3)}{\sqrt{3}}{\displaystyle \sum_{n=0}^\infty} c_n \lp\frac{1}{2}-\frac{\sigma_2}{\sigma_2+\sigma_3}\rp^{2n} \,\,  \mbox{if $\sigma_2 > 0$}
\end{array} \right .
\end{eqnarray}
where
\begin{equation}
\label{eq: A.16}
c_n=\sum_{k=0}^{2n}\frac{(4n\!-\!2k\!-\!5)!!(2k\!-\!1)!!(2n\!-\!1)!!}{2^n3^k(2n-k)!k!n!}\nn \,.
\end{equation}
This is the final result for incompressible flows in three spatial dimensions.

For the particular case (\ref{eq: 3.3})  considered by Smoluchowski \cite{Smo17} we have
\begin{equation}
\label{eq: A.17}
\ma A_{()} =
\left(\begin{array}{ccc}
0& 0&\alpha/2\\
0&0 & 0\\
\alpha/2&0 &0
\end{array} \right)\,.
\end{equation}
with eigenvalues $-\alpha/2,0,\alpha/2$. Substituting $\sigma_2 = 0$ and $\sigma_3 = \alpha/2$
into (\ref{eq: A.15}) we obtain (\ref{eq: 3.4}).
When $\sigma_2=0$ this expression agrees with the
classical result (\ref{eq: 3.4}) due to Smoluchowski \cite{Smo17}.


\begin{thebibliography}{}

\bibitem{Sha03} R. A. Shaw, {\it Annu. Rev. Fluid Mech.}, {\bf 35}, 183, (2003).

\bibitem{Wil08} M. Wilkinson, B. Mehlig, and V. Uski, {\it  Astrophys. J. Suppl.}, in press, (2008).

\bibitem{Smo17} M. v. Smoluchowski, {\it Zeitschrift f. physik. Chemie}, {\bf XCII}, 129-168, (1917); see eq. (28) on p. 156.

\bibitem{Saf56} P. G. Saffman and J. S. Turner, {\it J. Fluid Mech.}, {\bf 1}, 16-30, (1956).

\bibitem{Fal02} G. Falkovich, A. Fouxon and G. Stepanov, {\it Nature}, {\bf 419}, 151-154, (2002).

\bibitem{Wil06} M. Wilkinson, B. Mehlig and V. Bezuglyy, {\it Phys. Rev. Lett.}, {\bf 97}, 048501, (2006).

\bibitem{Bal01} E. Balkovsky, G. Falkovich, and A. Fouxon, {\it Phys. Rev. Lett.}, {\bf 86}, 2790, (2001); cond-mat/9912027.

\bibitem{Fal01} G. Falkovich, K. Gawedzki, and M. Vergassola, {\it Rev. Mod. Phys.}, {\bf 73}, 913, (2001).

\bibitem{Wil07} M. Wilkinson, B. Mehlig, S. \"Ostlund and K. P. Duncan,
{\it Phys. Fluids}, {\bf 19}, 113303, (2007).

\bibitem{And07} B. Andersson, K. Gustavsson, B. Mehlig, and M. Wilkinson, {\it Europhys. Lett.}, {\bf 80}, 69001, (2007).

\bibitem{Cre04} J. R. Cressman, J. Davoudi, W. I. Goldberg and J. Schumacher, {\it New J. Phys.}, {\bf 6}, 53, (2004).

\bibitem{Fal05} Falkovich {\em et al.}, {\it Nature}, {\bf 435}, 1045, (2005).

\bibitem{Fri97} U. Frisch, {\sl Turbulence}, Cambridge University Press, (1997).

\bibitem{Stu02} H. Sigurgeirson and A. M. Stuart, {\it Phys. Fluids}, {\bf 14}, 4352, (2002).

\bibitem{Kal07} J. Kalda, {\it Phys. Rev. Lett.}, {\bf 98}, 064501, (2007).

\bibitem{Som93} J. Sommerer and E. Ott, {\it Science}, {\bf 359}, 334, (1993).

\bibitem{Gra83} P. Grassberger and I. Procaccia, {\it Physica D}, {\bf 9}, 189, (1983).

\bibitem{Wil03} M. Wilkinson and B. Mehlig, {\it Phys. Rev. E}, {\bf 68}, 040101(R), (2003).





\end{thebibliography}
\end{document}